\documentclass[review,accept,moreauthors,pdftex,10pt,a4paper,preprints]{mdpi} 

\firstpage{1} 

\history{}

\usepackage[utf8]{inputenc}
\usepackage[T1]{fontenc}
\usepackage{hyperref}
\usepackage{graphicx,bm,amsmath,color,scalerel}
\usepackage{bbold}
\usepackage{wasysym}
\usepackage{verbatim}
\usepackage{physics}
\usepackage{comment}
\usepackage{subcaption}
\graphicspath{ {./img/} }
\usepackage{nomencl}
\makenomenclature
\usepackage{etoolbox}
\usepackage{cancel}
\renewcommand\nomgroup[1]{%
  \item[\bfseries
  \ifstrequal{#1}{Z}{Abbreviations}{%
  \ifstrequal{#1}{N}{Number Sets}{%
  \ifstrequal{#1}{O}{Other Symbols}{}}}%
]}

\newcommand{\be}{\begin{equation}}
\newcommand{\ee}{\end{equation}}
\newcommand{\beq}{\begin{eqnarray}}
\newcommand{\eeq}{\end{eqnarray}}
\newcommand{\ba}{\begin{align}}
\newcommand{\ea}{\end{align}}

\Title{Cosmic neutrinos as a window to departures from special relativity}

\Author{José Manuel Carmona$^{1}$*\orcidA{}, José Luis Cortés$^{1}$\orcidB{}, José Javier Relancio$^{1,2}$\orcidC{} and Maykoll A. Reyes$^{1}$\orcidD{}}

\AuthorNames{José Manuel Carmona, José Luis Cortés, José Javier Relancio and Maykoll A. Reyes}

\address{%
$^{1}$ \quad Departamento de F\'{\i}sica Te\'orica and Centro de Astropartículas y Física de Altas Energías (CAPA),
Universidad de Zaragoza, Zaragoza 50009, Spain; jcarmona@unizar.es (J.M.C); cortes@unizar.es (J.L.C.); mkreyes@unizar.es (M.A.R.)\\
$^{2}$  \quad Departamento de Física, Universidad de Burgos, 09001 Burgos, Spain; jjrelancio@ubu.es (J.J.R.)\\
}

\corres{Correspondence: jcarmona@unizar.es}

\abstract{We review the peculiarities that make neutrino a very special cosmic messenger in high-energy astrophysics, and, in particular, to provide possible indications of deviations from special relativity, as it is suggested theoretically by quantum gravity models. In this respect, we examine the effects that one could expect in the production, propagation, and detection of neutrinos, not only in the well-studied scenario of Lorentz Invariance Violation, but also in models which maintain, but deform, the relativity principle, such as those considered in the framework of Doubly Special Relativity. We discuss the challenges and the promising future prospects offered by this phenomenological window to physics beyond special relativity.}


\begin{document}

\section{Introduction}
\label{sec:introduction}

Neutrinos are very special elementary particles in the Standard Model (SM) of particle physics. Its existence was proposed for the first time~\cite{Pauli:1930pc}, not on the basis of a direct observation, but to solve an apparent paradox (the lack of energy balance in some nuclear decays). In fact, the approach used for neutrinos has served recently as a guide to several proposals to extend the SM based on the addition of very light and weekly coupled new (not directly observed) particles, introduced to solve some theoretical problems (QCD axion~\cite{Peccei:1977hh,Peccei:1977ur,Weinberg:1977ma,Wilczek:1977pj,Kim:1979if,Shifman:1979if,Dine:1981rt}) or to identify candidates for the particle content of dark matter (axion like particles~\cite{Preskill:1982cy,Abbott:1982af,Dine:1982ah}, feebly interacting particles~\cite{Batell:2009di,Beacham:2019nyx,Arguelles:2019xgp,EuropeanStrategyforParticlePhysicsPreparatoryGroup:2019qin,Lanfranchi:2020crw}). Another peculiarity of neutrinos is that, although originally thought as massless particles, we now know (through the observation of the phenomenon of neutrino oscillations~\cite{Fantini:2018itu}) that they have very small mass differences. We still do not know the absolute mass, nor whether they are Majorana or Dirac fermions~\cite{Bilenky:2020wjn}.   

Neutrinos are also very special messengers in high-energy astrophysics. In conventional physics, they are only affected in their propagation by the expansion of the Universe, and then, their observation gives us information of their production. Cosmic neutrinos are produced by the disintegration of pions or nuclei which are created by strong interactions, inside or close to astrophysical accelerators, or alternatively, by the decays produced after the interaction of Ultra High Energy Cosmic Rays (UHECR), propagating from the astrophysical Very High Energy (VHE) accelerators to the Earth, with the electromagnetic background (cosmogenic neutrinos). The observed cosmic neutrino flux can then be used to get information about astrophysical accelerators, the evolution of the Universe, and the cosmological electromagnetic background. This direct information provided by the neutrinos can be contrasted with the one obtained from the observed flux of UHECR, whose relation with the flux emitted by the astrophysical accelerators is affected by the interactions produced in their propagation and also by the intergalactic magnetic fields, limiting the possibility to localize their sources. The other possible messengers, VHE $\gamma$-rays, suffer also interactions with the electromagnetic background, which limits the information we can extract from their detected fluxes.    

The observation of the cosmic neutrino flux gives us, therefore, information about different processes in an energy domain several orders of magnitude beyond the highest energies achieved in terrestrial accelerators. This can give access for the first time to new physics whose observable effects increase with the energy. An example could be the deviations from Special Relativity (SR), appearing as a trace of a quantum theory of gravity. Such a theory will require to go beyond the classical notion of spacetime which SR theories are based on. Finding inconsistencies, within the framework of SR theories, in the description of the different observations from the messengers in high-energy astrophysics, could then be one way to identify signals of a theory of quantum gravity~\cite{Addazi:2021xuf}.   

In this work, our aim is to present the different ingredients in the study of possible signals of a departure from SR in the observations of high-energy cosmic neutrinos. In Sec.~\ref{sec:BSR} we present the different ways to describe a deviation from the kinematics of SR. There are several recent reviews on this subject~\cite{Arguelles:2021kjg,Stecker:2022tzd} which partly overlap with this work. The main motivation for a new review is to include in the discussion the possibility of a departure from SR while maintaining the relativistic principle. The different mechanisms of production of cosmic neutrinos and the modifications due to a departure from SR are the content of Sec.~\ref{sec:production}. Sec.~\ref{sec:propagation} is dedicated to the study of the effects in the propagation of neutrinos that can affect the detected flux of cosmic neutrinos. Some aspects of this detection are the content of Sec.~\ref{sec:detection}. We conclude with a discussion of the present situation and future prospects of the use of high-energy cosmic neutrinos as a window to physics beyond SR in Sec.~\ref{sec:discussion}.   

\section{Beyond special relativity}
\label{sec:BSR}

There are two different ways in which one can go beyond the kinematics of SR.  One can consider adding to the SM Lagrangian new terms that violate Lorentz Invariance (LIV). In case one wants to preserve the  relativistic invariance, one should modify the transformations between inertial frames and accordingly modify the SR kinematics; this is what is called Doubly/Deformed Special Relativity (DSR). In the following we will apply both scenarios to the study of phenomenological traces of new physics in the observations of very high-energy cosmic neutrinos.

\subsection{Lorentz Invariance Violation}

A deviation from SR whose effects increase with the energy, which is the physics we claim one can explore through the observations of high-energy cosmic neutrinos, can be incorporated in the framework of Effective Field Theory (EFT). This is done by adding to the fields and symmetries that define the SM of particle physics, terms of dimension higher than four which are not 
invariant under boosts (neglecting a possible deviation in the rotational symmetry). This is known as the Standard Model Extension (SME)~\cite{Colladay:1998fq,Kostelecky:2008ts}. 

The most important effect of this extension is contained in the free part of the Lagrangian density, i.e. in the part which is quadratic in fields. This leads to a modification of the SR energy-momentum relation of a free particle (modified dispersion relation) 
\be
E \,\approx\, p + \frac{m^2}{2\,p} + \alpha\, \frac{p^{n+1}}{\Lambda^n} \quad\text{when}\quad m\ll p\ll\Lambda\,,
\label{E(p)}
\ee
where $E$, $p$ are the energy and the modulus of the momentum of a particle with mass $m$, $\Lambda$ and $n$ are the energy scale and order of correction which parametrize the deviations from SR, and $\alpha$ is a dimensionless constant which parametrizes the particle dependence of the LIV effects.
Different values of $\alpha$ for different neutrino mass eigenstates would affect the patterns of neutrino oscillations~\cite{Barenboim:2022rqu}. There exist very strong constraints to such effects~\cite{Stecker:2022tzd}. In this review we consider the same $\alpha$ for all neutrinos.

The dimension of the first quadratic term in the SME which violates Lorentz invariance is $D=(4+n)$. There are two alternative values, $n=1$ (linear case) or $n=2$ (quadratic case) considered in the studies of LIV. It is also possible to consider a minimal SME~\cite{Colladay:1996iz}, where there are only operators of dimension four or less. However, in this case the effects do not increase with the energy; since we are interested in the phenomenology of cosmic neutrinos, along this review we will limit ourselves to the linear and quadratic cases. 

A modified dispersion relation implies a modification of the expression of the velocity of a particle in terms of the energy which can lead to observable consequences from transient astrophysical phenomena (energy-dependent photon time delays), even if the energies of the observed particles are much smaller than the energy scale parametrizing the LIV. Another observable consequence of modified dispersion relations is the modification of the SR kinematics in the different particle processes which are relevant in high-energy astrophysics. The thresholds and the separation of kinematically allowed/forbidden processes (with respect to SR) are affected by the modified energy-momentum relation when the mass dependent and the LIV terms in (\ref{E(p)}) become comparable, i.e., when $(m^2/E^2)\sim (E/\Lambda)^n$~\cite{Mattingly:2005re,Liberati:2013xla,Addazi:2021xuf}. This happens for $E\ll \Lambda$, and then, one can have observable consequences of the LIV in high-energy astrophysics at energies much lower that the energy scale of LIV. 

An important issue in the discussion of the use of cosmic neutrinos as a window to LIV is the particle dependence of the departures from SR. If the modification of the dispersion relation comes directly from a quadratic term in the SME, then the gauge symmetry requires the dimensionless parameter $\alpha$ to be the same for charged leptons and neutrinos. In this case one can use the bounds on the scale $\Lambda$ from the absence of signals of LIV in observations involving electrons, to exclude the possibility that LIV effects will be discovered from high-energy cosmic neutrino observations. There is an alternative where the quadratic term responsible of the modification of the dispersion relation comes from a higher dimensional term involving the scalar field doublet responsible of the Higgs mechanism. In this case it is possible to get a modified neutrino dispersion relation with no modification in the electron energy-momentum relation while maintaining the gauge symmetry~\cite{Carmona:2012tp,Crivellin:2020oov}. 

\subsection{Doubly Special Relativity}

We have seen that considering a LIV scenario entails a loss of the relativity principle and the acceptance of a preferred reference frame, which is usually identified with the one defined by the homogeneity and isotropy of the  Cosmic Microwave Background (CMB). 
If one wants to maintain a relativity principle when going beyond SR, one has to consider  a deformation of the transformations relating the inertial reference frames. The deformation of SR, usually called DSR~\cite{AmelinoCamelia:2008qg}, is assumed to be parametrized by a new energy scale $\Lambda$ which does not usually affect the rotational symmetry, as in the case of LIV.

A necessary ingredient of this departure from SR at the kinematical level is a nontrivial characterization of a multi-particle system with a total energy and momentum differing from the sum of the energies and momenta of the particles. One then has a composition of energy and momentum which is non-symmetric under the exchange of the particles~\cite{Kowalski-Glikman:2002oyi}. One arrives to this conclusion from different perspectives of DSR. The starting point of this proposal is the attempt to make compatible the relativistic invariance with the presence of a minimal length~\cite{AmelinoCamelia:2000ge,AmelinoCamelia:2000mn}, which seems to be a characteristic of a quantum theory which incorporates consistently the gravitational interaction~\cite{Kato:1990bd,Susskind:1993ki,Garay1995,Hossenfelder:2012jw}. Such minimal length can be understood as a consequence of a non-commutativity in a generalization of the classical spacetime which requires to go beyond the usual implementation of continuous symmetries by Lie algebras. The new algebraic structure is a Hopf algebra~\cite{Majid:1995qg} with a non-trivial co-product which leads to a deformed kinematics with a non-symmetric composition of momenta~\cite{Majid1994,Lukierski:1992dt}. An alternative way to arrive at the same conclusion is to identify the non-commutativity of spacetime with a non-commutativity of translations in a curved momentum space which can also be related with a composition of momenta~\cite{Kowalski-Glikman:2002oyi,Carmona:2019fwf,Lizzi:2020tci,Carmona:2021gbg,Relancio:2021ahm}. This composition law is therefore a crucial ingredient differentiating DSR and LIV.

Together with the non-linear composition of momenta, the invariance under deformed Lorentz transformations will lead in many cases to a modification of the dispersion relation. As a consequence, in the kinematic analysis of a process in DSR one has to consider both a possible modification of the energy-momentum relation of the particles participating in it and a modification of the energy-momentum conservation law. The compatibility with the relativity principle, in comparison with the case of LIV, can be shown to produce a cancellation of the effects of the two modifications. Therefore, in order to have an observable consequence of the deformation of the kinematics in a process, one has to consider energies comparable to the energy scale $\Lambda$ of the deformation~\cite{Carmona:2014aba,Albalate:2018kcf,Relancio:2020mpa,Carmona:2020whi,Carmona:2021lxr}. This means that in order to have a signal of DSR in the particle processes which are relevant in high-energy astrophysics, and in particular, in the observations of cosmic neutrinos, it is necessary to consider an energy scale parametrizing the deformation of the kinematics of the order of the energy involved in those processes. At the same time, many of the constraints to the high-energy scale in the case of LIV, do not apply in the DSR scenario. 

The two previous kinematic ingredients of DSR rise several problems and apparent contradictions on the physical interpretation of the theory. On the one hand, a modification of the composition of momenta in a particle system (independently of the distance between the particles) implies a departure from the notion of absolute locality in spacetime~\cite{AmelinoCamelia:2011bm,AmelinoCamelia:2011pe}. The corresponding loss of the crucial property of cluster~\cite{Carmona:2019oph}, which is at the basis of the formulation of special-relativistic quantum field theory, originates the so-called spectator problem~\cite{Carmona:2011wc,Amelino-Camelia:2011gae,Gubitosi:2019ymi}. On the other hand, 
a modification of the dispersion relation with the associated modification of the velocity of a particle rises an apparent inconsistency of DSR when one applies the deformed kinematics to any system, including a macroscopic system (soccer ball problem~\cite{Hossenfelder:2007fy,Amelino-Camelia:2011dwc}). 

A way out to these difficulties is to assume that the deformation of the kinematics only applies to particles closer than a certain length scale. If the mean distance among constituents of a macroscopic system is larger than this length scale, then the deformed kinematics will not affect these systems. Also, particles which are not sufficiently close to the particles participating in an interaction will not affect the kinematics of this process. This assumption would imply the absence of certain effects in the propagation of free particles, such as energy-dependent time delays, allowing one to consider deformations of SR at energies much lower than the Planck scale. As we will notice later, these ideas affect the study of the possible effects of DSR in high-energy astrophysics. However, the identification of a consistent theoretical framework incorporating them is still an open question. 

\section{Production of high-energy cosmic neutrinos}
\label{sec:production}

The first place to look for possible effects of a departure from SR in the physics of cosmic neutrinos are the processes responsible for their production.   

Charged pions, produced in strong interactions inside or close to a very high-energy astrophysical accelerator, will decay producing very high-energy neutrinos. There are difficulties in identifying the accelerators producing the highest cosmic rays and the mechanism of acceleration. From our experience at terrestrial accelerators, one can expect a very large number of pions produced in each interaction of a UHECR with the gas or electromagnetic halos surrounding the accelerator. It is then very difficult to give a reliable estimate of the production of charged pions at an astrophysical accelerator before considering a possible modification due to a departure of SR. In the case of LIV, such departure  would require to consider a modification of SR kinematics in a process with a large number of particles with modified energy-momentum relations, which could be particle dependent. The processes producing the pions involve composite particles (protons, nuclei, pions) and, according to the ideas proposed in the previous section to avoid the inconsistencies of a physical interpretation of DSR, one would expect no effect of DSR in the production of charged pions (assuming that the size of pions as composite systems is much greater than the characteristic length scale at which new effects should appear). The same arguments can be applied to the production of neutrons or unstable nuclei that are also a source of cosmic neutrinos through beta decay. 

Together with the production of neutrinos associated with the strong interactions inside or close to a very high-energy astrophysical accelerator, one can also consider the production of neutrinos due to the interaction of UHECR propagating in the electromagnetic cosmic background (CMB and Extragalactic Background Light (EBL)). Those neutrinos are referred as cosmogenic neutrinos and produce a diffuse flux. The origin of the neutrinos is once more the decay of charged pions, neutrons, or unstable nuclei.   

These  decays are processes whose kinematics including the effects of LIV~\cite{Martinez-Huerta:2020cut} or DSR~\cite{Carmona:2021lxr} can be systematically studied. In the SR case, the standard way to proceed is to consider the decay in the rest frame of the decaying particle, and then go to the laboratory frame by applying a boost with a very large Lorentz factor corresponding to the very high energy of the decaying particle. But this strategy relies on the relativistic invariance, so that it cannot be followed in the case of LIV, and in the case of DSR, its use is also problematic due to the complex form of the deformed Lorentz transformation of a multi-particle system~\cite{Kowalski-Glikman:2002oyi,Carmona:2012un}.

A modification of the energy-momentum relation will affect the decays producing the neutrinos in different ways depending on the particle dependence of the LIV effects.
The simplest scenario considers a modification of the energy-momentum relation (\ref{E(p)}) only for neutrinos, neglecting the modification of the energy-momentum relation for other particles because the corresponding dimensionless parameters $\alpha$ are much smaller than those of neutrinos.

One can consider two cases depending on the sign of the dimensionless parameter $\alpha$. When $\alpha > 0$ (superluminal case), the energy of a neutrino with a given momentum will be higher than in the case of SR. This will affect the production of neutrinos in the decays of charged pions, in the subsequent decay of muons produced in the decay of charged pions or in the decay of neutrons or unstable nuclei. LIV will suppress the production of neutrinos as the energy increases and pions will become stable particles for energies larger than a critical value. If $\alpha<0$ (subluminal case), one has the opposite situation, with neutrinos having a lower energy than in the case of SR for a given momentum, and one will have an enhancement  in the production of neutrinos as the energy increases.

While in the case $n=2$ neutrinos and antineutrinos can both be either superluminal or subluminal simultaneously, the $n=1$ case comes from a CPT-odd term in the SME Lagrangian, and one will have $\alpha$ coefficients with opposite sign for neutrinos and antineutrinos. In this case, one has a more complex situation with a superposition of enhancements and suppressions in the production of neutrinos and antineutrinos. 

A systematic procedure that can be followed, in order to identify the effects of a departure from SR in the decays producing neutrinos, is to consider directly the decays in the laboratory system using the collinear approximation, where one makes an expansion in powers of the ratios of masses and the energy of the decaying particle and in the relative angles of the momenta of the particles. 
This strategy can be applied in both cases, LIV and DSR~\cite{Carmona:2022future}. In the latter case, one will have to include a modification of the total energy-momentum of the charged lepton-neutrino pair produced in the different decays responsible of the production of very high-energy cosmic neutrinos. DSR modifies in general the neutrino spectrum but, in contrast to the superluminal case of LIV, pions cannot become stable at any energy because of the relativistic principle~\cite{Amelino-Camelia:2011gae}.

\section{Propagation of cosmic neutrinos}
\label{sec:propagation}

The absence of interactions in the propagation of cosmic neutrinos is what makes them very special astrophysical messengers. In the case of SR kinematics, all the modification in their spectrum from their production until their detection at Earth is due to the expansion of the Universe.  

In the LIV scenario, a modification of the energy-momentum relation \eqref{E(p)} of neutrinos can affect the propagation of cosmic neutrinos. When $\alpha>0$, neutrinos become unstable and they can decay producing a pair of charged leptons~\cite{Carmona:2012tp,Carmona:2019xxp} or a neutrino-antineutrino pair~\cite{Mattingly:2009jf,Stecker:2014oxa}, together with a neutrino of the original flavor and lower energy. 

In the case of the production of an electron-positron pair (Vacuum Pair Emission or VPE), $\nu_i\to\nu_i + e^+ +  e^-$, the decay is allowed for energies above a given value (threshold), which is fixed by the electron mass and the scale $\Lambda$ of LIV. The decay width is proportional to $E^{5+3n}$~\cite{Carmona:2012tp}  and then the decay length is very small for energies above the threshold. The production of the electron-positron pair is accompanied by a loss of energy of the neutrino leading to a drastic suppression of the flux of neutrinos at Earth at energies above the threshold. 

In the scenario corresponding to $n=2$, one can have two cases depending on the sign of $\alpha$, so that in the case of $\alpha>0$ one can put a lower bound on the energy scale $\Lambda$ of LIV~\cite{Stecker:2014oxa,Carmona:2019xxp} from the observation of cosmic neutrinos beyond the PeV~\cite{Aartsen:2013jdh,IceCube:2013cdw,Aartsen:2014gkd,Aartsen:2017ibm,Aartsen:2018fqi}. In the case $\alpha<0$ the spectrum of the detected neutrinos is not modified by effects due to LIV in their propagation. 

In the case of a linear ($n=1$) modification of the energy-momentum relation, one has opposite signs of $\alpha$ for neutrinos and antineutrinos, and then one would have a drastic suppression in the flux of either neutrinos or antineutrinos. The possibility to detect the two fluxes separately could allow to identify this signal of LIV. At present, IceCube cannot distinguish between neutrino and antineutrino events; a signal from the $n=1$ case could still be present in the detected spectrum depending on the neutrino-antineutrino composition of the initial flux emitted by the source~\cite{Stecker:2014oxa}.

Together with the production of an electron-positron pair one can also have, in the case $\alpha>0$, the production of a neutrino-antineutrino pair (Neutrino Splitting or NSpl), $\nu_i\to \nu_i+\nu_j+\bar{\nu}_j$, where $i$ and $j$ are flavor indices that can be equal or not. This decay produces a loss of energy and a multiplication of neutrinos. In contrast to the production of an electron-positron pair, there is no threshold when one neglects the neutrino masses. However, looking at the decay width of the process one can identify a combination of constants (Fermi coupling $G_F$, Hubble constant $H_0$, and the energy scale of LIV $\Lambda$) that behaves as an effective threshold, energy below which one can neglect the energy loss and above which the decay length is very small, similarly to the VPE process~\cite{Carmona:2019xxp}.
The decay width is also proportional to $E^{5+3n}$; then, the observation of cosmic neutrinos constrains the simplest scenario with $n=2$ and $\alpha>0$. 
In order to arrive at the same conclusion in the linear case, it is necessary to identify a flux of both cosmic neutrinos and antineutrinos~\cite{Stecker:2014oxa}. 

In the case of DSR, the compatibility of the deformed kinematics with the (deformed) relativistic invariance forbids the decay of neutrinos, and then, there cannot be a signal of DSR in the propagation of neutrinos. Strictly speaking, one can have a decay of a neutrino into two neutrinos and an antineutrino of smaller masses. This decay is also possible in SR, although the lifetime is several orders of magnitude bigger than the lifetime of the Universe, so that in practice one can consider neutrinos as stable particles. The same result applies to DSR, as one can see if one uses the collinear approximation~\cite{Carmona:2022future} to the decay of neutrinos.  

\section{Detection of cosmic neutrinos}
\label{sec:detection}
\medskip

Charged current interactions create charged leptons which produce  signals through Cherenkov radiation in neutrino detectors. The use of ice in the South Pole as a detector (IceCube) has been the first experiment to detect high-energy cosmic neutrinos in the ($100\, \text{TeV}$--$10\, \text{PeV}$) energy range. The observed flux can be used to put limits on (or identify hints of) departures from SR. For example, Ref.~\cite{Stecker:2014oxa} explored the possibility that the apparent end of the cosmic neutrino spectrum~\cite{Ahlers:2013xia} could be due to a violation of Lorentz invariance. The detection~\cite{IceCube:2021rpz} of an event compatible with the Glashow resonance (an enhancement in the probability of interaction of electron antineutrinos with electrons of the ice due to the production of a real $W$~\cite{Glashow:1960zz}), however, diminishes the indications for a cutoff in the neutrino spectrum.

In the near future, there are several proposals (see last section of Ref.~\cite{Stecker:2022tzd}) to improve the sensitivity to high-energy cosmic neutrinos and measure the flux of cosmic neutrinos in the ($10\, \text{PeV}$--$ 1\, \text{EeV}$) energy range. Those experiments will enlarge the window that cosmic neutrinos open to explore departures from SR. 

The interactions used in the detection of extremely high-energy neutrinos can also be affected by the departures from SR and should be taken into account if one wants to find a signal of those possible deviations in the detected flux of cosmic neutrinos. In both cases, LIV and DSR, the kinematics of the interactions used in the detection of cosmic neutrinos will be affected. In the case of LIV, the effect is due to the modified energy-momentum relation of the neutrinos; in DSR, it is due to the modification of the composition of the momentum of the detected neutrino and the momentum of the produced charged lepton in a local interaction.  

\section{Discussion and future prospects}
\label{sec:discussion}
\medskip

We have presented a qualitative discussion of the different ingredients in the study of high-energy cosmic neutrinos where a departure from SR (either LIV or DSR) could lead to observable effects. In the case of LIV where one neglects all effects except for the modification of the energy-momentum relation for the neutrino, one could go easily beyond the qualitative discussion trying to extract the value of (or put limits on) the unique parameter $\alpha/\Lambda^n$, which controls all the effects of LIV. Beyond this especially simple case, tests of LIV would require to consider different scenarios depending on how the LIV affects the kinematics of all the particles involved in the production, propagation, and detection of cosmic neutrinos. In the case of DSR, one has different models depending on the choice of the modified composition law of energy and momentum. They involve the energy scale $\Lambda$ of the deformation as a unique parameter, but a quantitative analysis should be done for each model. Such analysis goes beyond the aim of the present review.

The main challenge of using the observations of cosmic neutrinos as a window to departures from SR is to be able to reduce the uncertainties on the understanding of the production of cosmic neutrinos, either in VHE astrophysical accelerators or in the propagation of UHECR. Any progress in astrophysics about the mechanisms of acceleration of particles to the highest possible energies in astrophysical accelerators would be very important to improve the sensitivity of the observations of cosmic neutrinos to possible departures from SR. The same argument applies to a progress about the identification of the composition of UHECR (protons and heavier nuclei), and on the electromagnetic background through which UHECR propagate.    

A very special role in the attempts to make progress can be played by the use of the Multi Messenger (MM) strategy, comparing the VHE-UHE neutrino fluxes with the VHE, $\gamma$-ray flux, and the end of the  UHECR  detected flux. New experiments and proposals in the detection of the three mentioned messengers will surely improve the present situation about the study of signals of a departure from SR in high-energy astrophysics. A  complementary window in this MM strategy is the very active program dedicated to the detection of gravitational waves.

Almost all the studies of possible signals of a departure from SR in the high-energy cosmic neutrino flux have been concentrated until now on the case of LIV.
It is interesting to go one step further exploring the possible signals of a modification due to DSR with a sub-Planckian energy scale of deformation, to the different processes involved in the production and detection of high-energy cosmic neutrinos.
A systematic study of those modifications require a formulation of the collinear approximation~\cite{Carmona:2022future} in the processes involving very high-energy elementary particles at very small distances.

Besides the window to physics beyond SR provided by the observation of very high-energy cosmic neutrinos, one should keep an eye open to possible surprises from the future results on the very precise attempts to measure the mass of the neutrino~\cite{KATRIN:2021uub}. Any clue of a violation of SR coming from these attempts could alter our understanding of the role that neutrinos play in cosmology as the more abundant component of the Universe.

To summarize, we are in a situation where one can expect new results from recently open windows to observations of the different cosmic messengers which can be sensitive for the first time to physics beyond SR. The present review has focused on one of these new windows, very high-energy cosmic neutrinos, presenting the different ingredients involved in their study. We have discussed the present situation and some suggestions about how to proceed in the near future.

\section*{Acknowledgments}
This work is supported by Spanish grants PGC2018-095328-B-I00, funded by MCIN/AEI/10.13039/501100011033
and by ERDF A way of making Europe,  and DGA-FSE grant 2020-E21-17R. JJR acknowledges support from the Unión Europea-NextGenerationEU (``Ayudas Margarita Salas para la formación de jóvenes doctores'').  The work of MAR is supported by MICIU/AEI/FSE (FPI grant PRE2019-089024). This work has been partially supported by Agencia Estatal de Investigaci\'on (Spain)  under grant  PID2019-106802GB-I00/AEI/10.13039/501100011033.  The authors would like to acknowledge the contribution of the COST Action CA18108 ``Quantum gravity phenomenology in the multi-messenger approach''.



\end{document}